\documentclass[12pt]{article}

\usepackage{sbc-template}
\usepackage{graphicx,url}
\usepackage[utf8]{inputenc}
\usepackage[english]{babel}
\usepackage{subcaption}
\usepackage{xcolor}

\title{Adaptive Detection of Software Aging under Workload Shift}

\author{Rafael José Moura\inst{1}, Maria Gizele Nascimento\inst{1}, Fumio Machida\inst{2}, Ermeson Andrade\inst{1}
}

\address{Federal Rural University of Pernambuco (UFRPE) \\ Recife, PE -- Brazil
    \email{\{rafael.mourasilva, gizele.alves, ermeson.andrade\}@ufrpe.br}
\nextinstitute
    University of Tsukuba \\ Tsukuba -- Japan
    \email{machida@cs.tsukuba.ac.jp}
    }

\begin{document} 

\maketitle

\begin{abstract}
  Software aging is a phenomenon that affects long-running systems, leading to progressive performance degradation and increasing the risk of failures. To mitigate this problem, this work proposes an adaptive approach based on machine learning for software aging detection in environments subject to dynamic workload conditions. We evaluate and compare a static model with adaptive models that incorporate adaptive detectors, specifically the Drift Detection Method (DDM) and Adaptive Windowing (ADWIN), originally developed for concept drift scenarios and applied in this work to handle workload shifts. Experiments with simulated sudden, gradual, and recurring workload transitions show that static models suffer a notable performance drop when applied to unseen workload profiles, whereas the adaptive model with ADWIN maintains high accuracy, achieving an F1-Score above 0.93 in all analyzed scenarios.
\end{abstract}
     
\section{Introduction}

With the increasing reliance on complex and long-running software systems, such as database servers, cloud applications, and embedded systems, reliability and availability have become essential requirements to ensure the stability and efficient operation of these systems. One of the main phenomena that threaten this stability is software aging, which progressively compromises system performance and can lead to failures over time~\cite{cotroneo2014survey}. This problem is generally caused by faults such as memory leaks, resource fragmentation, and the accumulation of internal errors~\cite{jain2020detection, dabukke2025implementation}. When not properly identified and mitigated, software aging can result in severe failures, leading to service degradation and even financial losses for organizations that depend on software for critical operations.

To mitigate the effects of software aging, several strategies have been proposed, including software rejuvenation techniques. However, for these techniques to be effective, it is essential to accurately and proactively detect the signs of aging~\cite{pietrantuono2020survey}. In this context, Machine Learning (ML)-based approaches have proven to be promising due to their ability to identify degradation patterns from system monitoring data. Nevertheless, most traditional ML models assume that the statistical distribution of data remains stationary over time, which rarely occurs in practice. Real software systems operate in dynamic environments, with usage patterns and resource consumption that constantly change according to different workloads. Workload shifts can be understood as changes in the intensity of request rates per second sent to the server. Such changes directly affect resource consumption, and higher workloads accelerate the manifestation of software aging symptoms, including increased memory usage, response time degradation, and faster progression toward memory exhaustion~\cite{couto2024comparative}. These changes directly impact the performance of predictive models, as they alter the relationship between the input variables and the expected output.

This work aims to evaluate the robustness and adaptability of software aging detection models in environments with varying workload conditions. Rather than addressing concept drift in a strict sense, we analyze how static and adaptive models respond to explicit workload shifts that alter system behavior and affect model performance. We investigate scenarios involving sudden, gradual, and recurring workload transitions, comparing a static baseline model with adaptive models that integrate adaptive detectors such as the Drift Detection Method (DDM) and Adaptive Windowing (ADWIN). Although originally developed for concept drift detection in data streams, these techniques are adapted in this study to address performance loss caused by workload shift, demonstrating their potential to support adaptation in dynamic operational environments. Experimental results show that, under such workload shifts, static models suffer significant performance degradation, while the adaptive approach, especially with ADWIN, maintains high detection accuracy.

The remainder of this paper is organized as follows: Section~\ref{sec:teorics} presents the theoretical background, including the concept of software aging and adaptive detectors, detailing the DDM and ADWIN. Section~\ref{sec:works} discusses related work. Section~\ref{sec:method} describes the methodology employed. Section~\ref{sec:results} presents the experiments conducted and the analysis of the results. Finally, Section~\ref{sec:conclusion} presents the conclusions of the study and outlines directions for future work.

\section{Background} \label{sec:teorics}

This section presents the main theoretical foundations that support this study. It discusses the phenomenon of software aging and introduces the adaptive detectors adopted, specifically DDM and ADWIN, and how these methods are applied to maintain the robustness of software aging detection models under varying workload conditions.

\subsection{Software Aging} \label{sec:Aging}

Software aging refers to the performance degradation or the emergence of failures after a system has been running continuously for an extended period, due to the accumulation of internal errors that lead to the exhaustion of computational resources~\cite{cotroneo2014survey}. This degradation occurs mainly because of the system's inadequacy to meet new environmental and user demands, as well as the cumulative modifications made over time~\cite{parnas1994software}. To mitigate these effects, software aging detection and software rejuvenation techniques are adopted. Software rejuvenation is a proactive approach that performs system state cleanup and periodic restart, aiming to prevent failures related to aging~\cite{jia2017software, cotroneo2014survey}.

Software aging detection is performed by monitoring system indicators such as available physical memory, RAM usage, swap space, file and process table sizes, response time, and traffic metrics, including packet and bit rates~\cite{Shruthi2020AnAO}. More recently, ML-based approaches have been successfully applied in this context, using the same indicators as input features to train models capable of classifying the system's state or predicting the failure moment due to resource exhaustion, such as memory \cite{nascimento2024comparison, jia2017software, cotroneo2014survey}. ML-based approaches are grouped into three categories: classification (identifying the system state,~\cite{andrzejak2008using}), regression (estimating time to failure (TTF)~\cite{nascimento2024comparison}), and time series forecasting (predicting future resource trends using models like ARIMA or Long Short-Term Memory (LSTM) \cite{carberry2024real}). However, these approaches generally assume that the data distribution remains constant over time. In real-world environments, this assumption rarely holds, as changes in system usage patterns can affect the relationship between monitored indicators and the system state.

\subsection{Adaptive Detectors: DDM and ADWIN} \label{sec:change-detection}

To address performance degradation caused by variations in workload, we incorporate two adaptive learning techniques: the DDM and ADWIN. Although originally developed for detecting distribution changes in data streams, these methods can also be used to trigger model updates in response to operational changes that affect prediction accuracy.

The DDM monitors the online error rate of a model and applies statistical bounds to detect significant increases. When the error rate first exceeds a warning threshold and then a drift threshold, the method signals a distributional change and triggers model adaptation~\cite{gama2004learning}. ADWIN, in turn, maintains a variable-length window of recent data and automatically detects shifts by splitting the window into two sub-windows and comparing their statistics. If the difference between them is statistically significant, older data are discarded and the model is updated with the most recent information~\cite{bifet2007learning}.

In this work, both methods are applied to controlled workload shift scenarios, such as abrupt, progressive, or periodic changes in load intensity, that can impact system behavior and reduce model generalization. As the system continues to operate over time, its failure behavior and resource consumption may evolve. Such variations can reduce the accuracy and effectiveness of ML-based detection models. By integrating DDM and ADWIN, we aim to handle abrupt or recurring workload shifts, enabling models to maintain robustness and adapt to changing operational conditions.

\section{Related Works} \label{sec:works}

The literature on software aging has mainly focused on detecting this phenomenon or predicting when it will occur. Several approaches have been proposed, including hybrid strategies that combine statistical techniques and ML to improve detection accuracy in different application contexts. However, most of these studies do not consider the impact of workload shift on the manifestation of software aging. To structure this discussion, we first present works that explicitly analyze the effects of workload shift, then describe studies that use ML techniques for aging detection, followed by approaches that investigate performance degradation caused by workload changes. Finally, we position our contribution in relation to these existing efforts.

Workload shift has been shown to directly influence software aging behavior. Watanabe et al.~[2023] analyze a real-time object detection system deployed on an edge server and demonstrate that varying the input image size leads to statistically significant degradation in memory and swap usage, resulting in reduced time-to-failure events. These findings highlight that changes in workload intensity can accelerate aging symptoms and negatively affect overall system reliability. Similar results are reported in Couto et al.~[2024] and Andrade et al.~[2021], which likewise show that workload fluctuations exacerbate aging-related degradation in long-running systems. \nocite{watanabe2023software}~\nocite{couto2024comparative}~\nocite{andrade2021comparative}

Several ML–based approaches have been proposed to detect aging in long-running systems. Battisti et al.~[2022] introduce hLSTM-Aging, a hybrid model combining a Convolutional Long Short-Term Memory (Conv-LSTM) network and a Moving Average (MA) model to forecast resource usage trends and anticipate failures caused by aging. Jia et al.~[2023] propose DGRU (Decomposition-based Gated Recurrent Unit), which leverages Seasonal-Trend Decomposition (STL) and GRU networks to model long-term growth and short-term fluctuations in memory usage series to predict aging-related degradation. Nie et al.~[2024] present MSAP (Multi-Scenario Aging Prediction), an ensemble learning approach for aging prediction on Android systems that uses multiple workload profiles and ML algorithms to maintain robustness in highly dynamic environments.\nocite{battisti2022hlstm}\nocite{jia2023software}\nocite{nie2024method}

In addition to aging detection, some studies have addressed performance degradation caused by variations in workload. For example, Huang et al.~[2023] analyze how changes in workload intensity and co-located Virtual Machine (VM) interference produce large execution-time increases, and propose an ML predictor and a performance-degradation index to distinguish workload-induced slowdowns from interference.  Similarly, Meyer et al.~[2021] show that variations in co-located application workloads create cross-application interference that significantly increases response time and leads to Service Level Agreement (SLA) violations, motivating interference-aware ML scheduling. Lastly, Ahmed et al.~[2023] report empirical evidence that CPU steal time leads to measurable performance degradation in cloud VMs under varying load conditions, and propose a detailed framework for monitoring the issue. \nocite{huang2024cloudprophet}\nocite{meyer2021ml}\nocite{ahmed2023exploring}

Despite these advances, to the best of our knowledge there are no studies evaluating how aging detection models perform under dynamic workload conditions. This work addresses this gap by comparing static and adaptive ML models in controlled workload shift scenarios and demonstrating how adaptive detectors, such as DDM and ADWIN, can help maintain detection accuracy in dynamic operational environments.

\section{Methodology} \label{sec:method}

\begin{figure}[h!]
    \centering
    \includegraphics[width=0.9\linewidth]{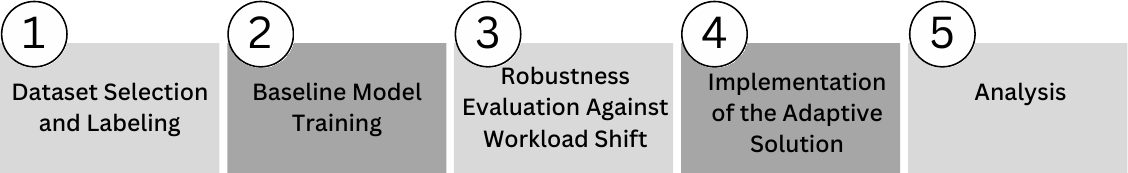}
    \caption{Stages of the methodology adopted in this study.}
    \label{fig:metodologia}
\end{figure}

The methodology adopted in this study is structured into five main steps, aiming to select, label, and detect software aging, with the objective of investigating the effects of workload shift on ML models for aging detection. Figure~\ref{fig:metodologia} presents a summary of the steps applied in this methodology, illustrating the process flow from data acquisition to the evaluation of adaptive models.


\subsection{Dataset Selection and Labeling}

To investigate the impact of workload shift on software aging detection, we used a dataset from the experimental study by~\cite{couto2024comparative}, which monitored RAM consumption in a SQL Server DBMS over 48 hours under different workload levels (``Low'', ``Medium'' and ``High'') with measurements taken every five seconds. Due to the scarcity of publicly labeled datasets for software aging, a custom labeling strategy based on raw monitoring data was developed. This approach aims to distinguish persistent degradation from natural fluctuations and random noise~\cite{machida2013effectiveness}. A binary labeling procedure (Normal/Aging) was applied individually to each workload profile, following four main steps:

\begin{itemize}
    \item \textbf{Removal of the {Warm-up} Phase:} Initially, the data were preprocessed to remove an initial {warm-up} phase, corresponding to the first 600 seconds of monitoring. This step is crucial to ensure that the subsequent analysis focuses on the system's stable behavior, disregarding transient variations during startup.

    \item \textbf{{STL decomposition}:} The memory time series, after removal of the {warm-up} phase, was processed using the STL decomposition technique. This method separates the original signal into three components: trend (long-term behavior), seasonality (repetitive patterns), and residual (random noise)~\cite{jia2023software}. The trend component is essential for aging detection, as it reflects the underlying growth or decline in memory consumption.


    \item \textbf{Trend Analysis and Labeling via Linear Regression in Sliding Windows:} The isolated trend component is analyzed using linear regression applied in sliding windows. For each window, the slope of the regression line is calculated. If it exceeds a fixed threshold of 0.5 memory units per sample, the window is labeled as {Aging} (1). This approach identifies continuous memory growth periods characteristic of software aging while remaining robust to short-term fluctuations. The threshold was chosen empirically to balance sensitivity and robustness: smaller values tend to misclassify short-term fluctuations as aging, whereas higher values may delay the detection of genuine long-term trends.
    
    \item \textbf{Binary Label Consolidation:} Finally, the complete time series is labeled in a binary manner. Windows identified with a significant growth trend are classified as {Aging} (1). The remaining parts of the series, including the {warm-up} phase (explicitly labeled 0) and periods without a detectable growth trend, are classified as Normal (0). This consolidation integrates the trend labeling with the {warm-up} phase, providing a comprehensive view of the system's behavior over time.
\end{itemize}

\subsection{Baseline Model Training}
\label{ssec:baseline_model}

With the data properly labeled, the next step is the construction of a baseline model. For this purpose, a supervised ML model based on {Random Forest} (RF) was adopted, trained under a single workload scenario. In this study, the ``Low'' workload was chosen in order to create a static environment that allows for measuring the model's optimal performance.

The objective of this phase is to establish the theoretical maximum performance of the model. The evaluation is conducted in the same workload context used for training, using standard classification metrics: Accuracy, Precision, Recall, and F1-Score. The results obtained define the performance baseline, which will be used as a reference to quantify the degradation under workload shift.

\subsection{Robustness Evaluation Against Workload Shift}

After training the reference model under a “Low” workload, we next assess its robustness when applied to different workload levels. Rather than focusing on abstract distributional drift, we simulate workload shifts that directly alter the system’s resource‐usage patterns:

\begin{itemize}
    \item \textbf{Sudden {Shift}}, characterized by an abrupt transition from one workload profile to another, such as a direct shift from ``Low'' to ``High'' workload; 
    \item \textbf{Gradual {Shift}}, representing a progressive transition with incremental mixing of data from different workloads, simulating a constantly evolving environment; 
    \item \textbf{Recurring {Shift}}, involving periodic alternation between distinct workload profiles, reflecting cyclic operational patterns of the system. 
\end{itemize}

For each scenario, we apply the reference model (trained on Low) to the shifted dataset and compute Accuracy, Precision, Recall, and F1‐Score. This analysis quantifies the model’s sensitivity to workload‐induced changes, with the goal of identifying conditions where adaptive strategies are required to maintain reliable software aging detection under variable loads.

\subsection{Implementation of the Adaptive Solution}

To mitigate the performance degradation caused by workload-induced shifts, we implement an adaptive solution based on continuous model retraining. This strategy combines statistical adaptive detectors with automatic update mechanisms, enabling the model to dynamically adjust to shifting input patterns. Two detectors from the \texttt{river} library~\cite{riverlib} are employed: DDM and ADWIN. Both monitor the model's error rate for each new instance and trigger retraining when significant shifts are detected according to statistical criteria.

In the case of DDM, upon detecting a change, the model is retrained using a sliding window with the last 2000 processed samples. Retraining is performed only when the window contains more than one class, to avoid imbalance. After the update, the detector is reset, initiating a new detection cycle. For ADWIN, the same mechanism is adopted: upon detecting a change, the model is retrained with the last 2000 instances in a sliding window. As with DDM, retraining occurs only if the window contains multiple classes. The detector is then reset to monitor the new operating condition. This approach enables the model to maintain its accuracy even in the presence of abrupt, gradual, or recurring workload changes, promoting greater predictive robustness in dynamic environments.

\subsection{Analysis}

The final step performs a systematic comparison between the static model (trained only with data from the Low workload) and the adaptive models integrated with DDM and ADWIN. The evaluation considers four distinct workload shift scenarios: two cases of sudden workload shift (transitions from Low to Medium and Low to High), one gradual workload shift, and one recurring workload shift. To ensure reproducibility and transparency, all the data used, as well as the trained models, 
are available in an online repository\footnote{\url{https://anonymous.4open.science/r/drift-aging-detection-E573/README.md}}. 
Furthermore, the obtained results are analyzed, aiming to identify the contexts in which each detector demonstrates greater effectiveness, as well as their operational limitations.

\section{Experiments and Results} \label{sec:results}

This section presents the experiments conducted to evaluate the effectiveness of the proposed approach. Initially, the training of the baseline model is described, followed by the {simulation of different types of workload shifts} and the application of adaptive detectors. The analysis of the results aims to compare the performance between static and adaptive models in scenarios with dynamic workload shifts.

\subsection{Model Training and Workload Shift Simulation}

To assess the effectiveness and robustness of the detection model, the experiments were divided into two stages. First, a baseline model, based on the RF algorithm, was trained and validated using {k-fold} cross-validation with $k=5$, exclusively with data from the Low workload scenario. The objective of this step was to establish a performance baseline in a static and controlled environment. The value of $k=5$ was chosen as it provides a stable evaluation of the model with a good balance between performance and computational cost.

Table~\ref{tab:desempenho_modelo} summarizes the performance obtained under different workload scenarios. It is observed that the model, trained only with data from the Low workload, does not generalize well to other contexts: the F1-Score, for example, dropped from 0.9996 to 0.7689 in the Medium workload and to 0.8207 in the High workload. This degradation highlights the impact of workload shift, indicating that a static model is insufficient to capture changes in the system's behavioral patterns.

\begin{table}[h!]
\caption{Model performance under different workload profiles}
\centering
\small
\renewcommand{\arraystretch}{1.1} 
\begin{tabular}{lcccc}
\hline
\textbf{Test Scenario} & \textbf{Accuracy} & \textbf{Precision} & \textbf{Recall} & \textbf{F1-Score} \\
\hline
Low Workload    & 0.9995 & 0.9996 & 0.9996 & 0.9996 \\
Medium Workload & 0.6542 & 0.7626 & 0.7753 & 0.7689 \\
High Workload   & 0.7356 & 0.8028 & 0.8395 & 0.8207 \\
\hline
\end{tabular}
\label{tab:desempenho_modelo}
\end{table}

\begin{figure}[h!]
    \centering
    \begin{subfigure}[b]{0.49\linewidth}
        \centering
        \includegraphics[width=\linewidth]{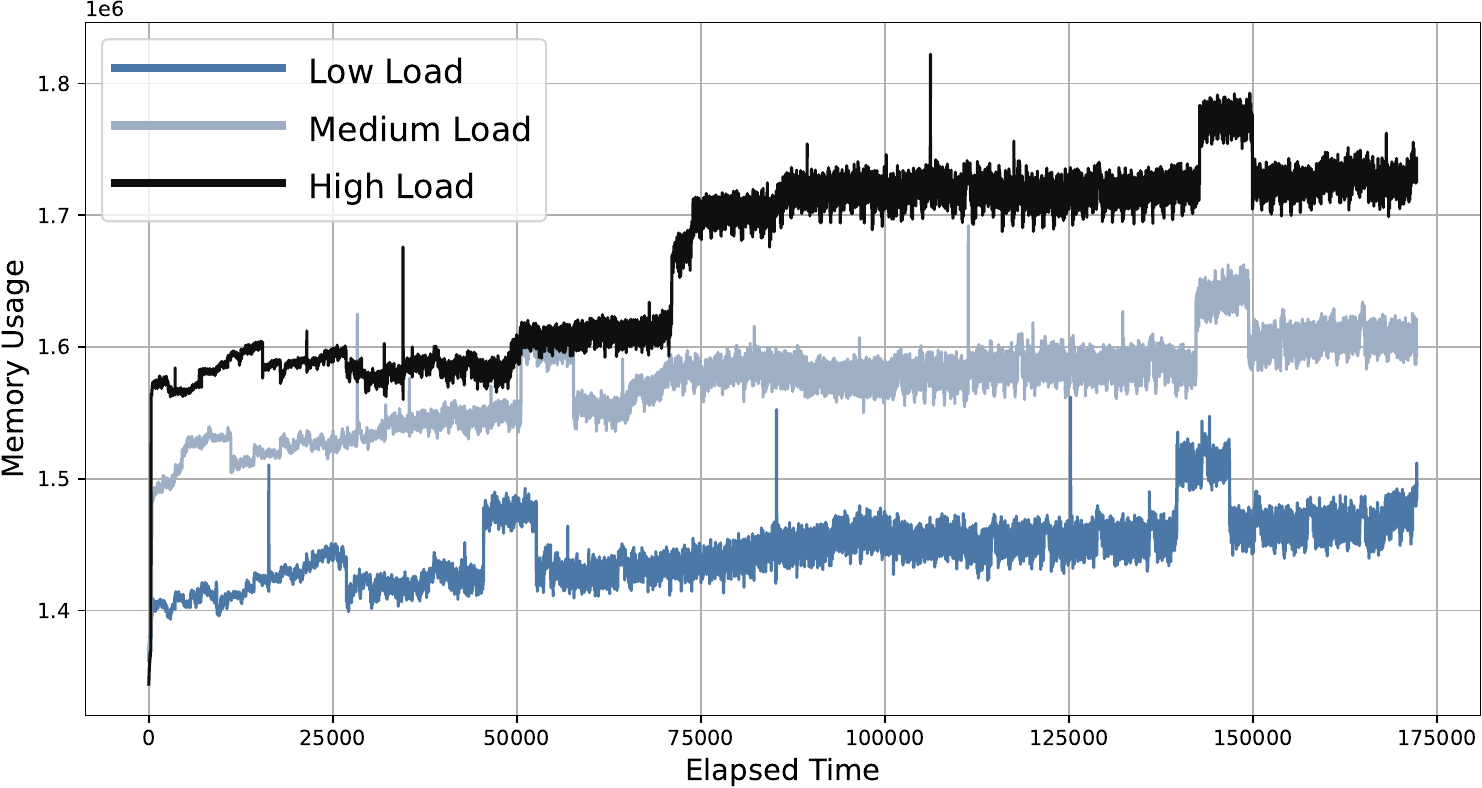}
        \caption{Memory consumption over time.}
        \label{fig:memoria}
    \end{subfigure}
    \hfill
    \begin{subfigure}[b]{0.49\linewidth}
        \centering
        \includegraphics[width=\linewidth]{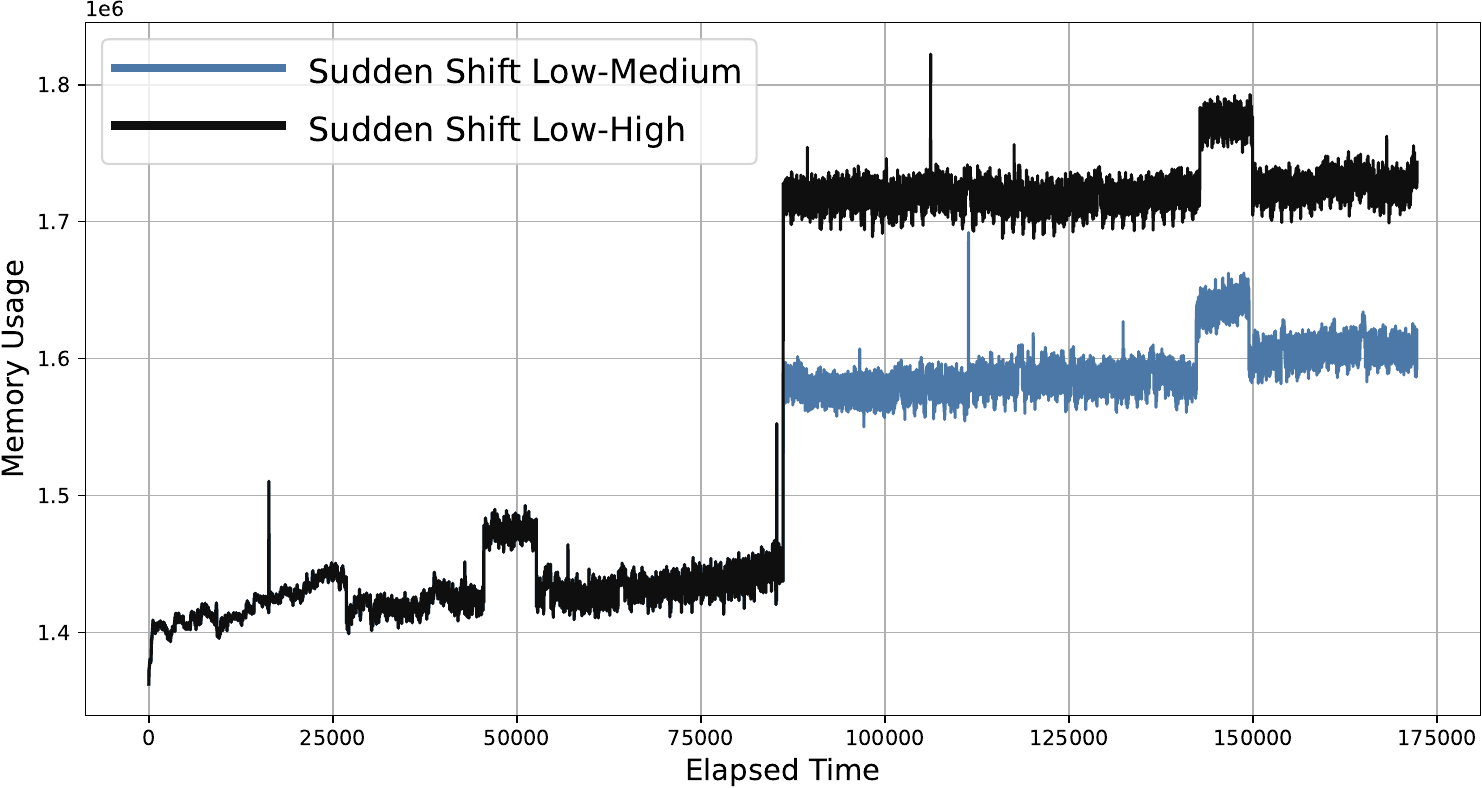}
        \caption{Sudden drift simulation.}
        \label{fig:memoria_drift_subito}
    \end{subfigure}

    \vspace{0.5cm}

    \begin{subfigure}[b]{0.49\linewidth}
        \centering
        \includegraphics[width=\linewidth]{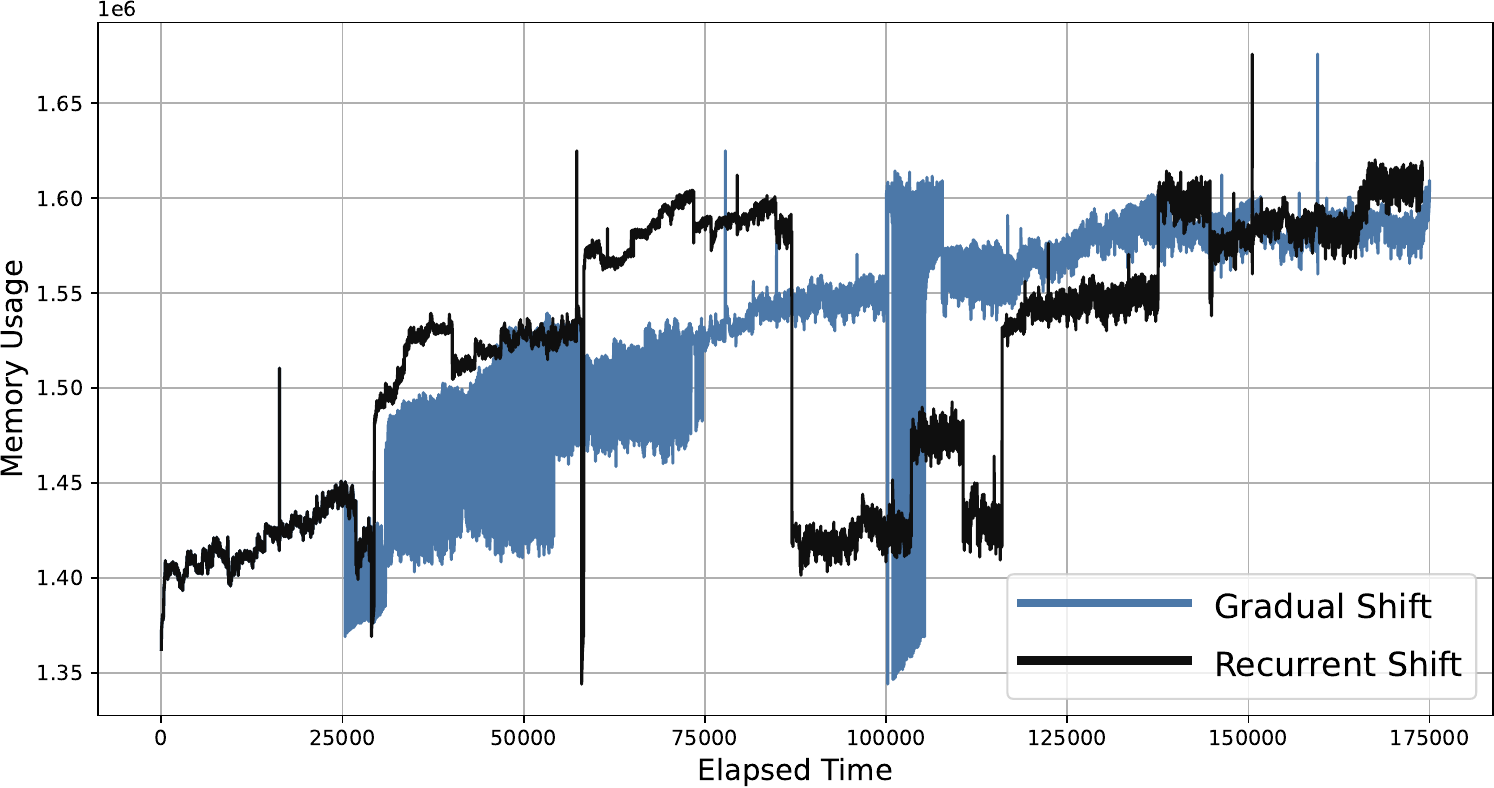}
        \caption{Gradual and recurring drift simulation.}
        \label{fig:memoria_drift_recorrente}
    \end{subfigure}
    \hfill

    \caption{Memory usage and drift scenarios.}
    \label{fig:memory_matrix}
\end{figure}

Figure~\ref{fig:memoria} illustrates the difference between the memory consumption scenarios for each workload, showing the distinct system usage dynamics under different levels of stress. This helps explain the model's difficulty in generalizing, as it is evident that each workload exhibits a different memory consumption behavior over time. To investigate the impact of more dynamic data changes, synthetic datasets were generated based on the experimental data presented in~\cite{couto2024comparative}, simulating different types of workload shift. Figure~\ref{fig:memoria_drift_subito} shows the sudden shift simulation, characterized by an abrupt transition between workload profiles. Two scenarios were considered: a shift from Low to Medium workload and another from Low to High workload. Figure~\ref{fig:memoria_drift_recorrente} illustrates the cases of gradual and recurring shift, simulated through the mixing and alternation of different workloads, representing a gradual evolution and a periodic repetition of patterns.

Table~\ref{tab:resultados_rf} presents the performance of the baseline model in different workload shift scenarios, considering multiple evaluation metrics. However, the following analysis focuses on the F1-Score, as it is the most appropriate metric to assess the extent to which adaptive models can recover or maintain performance in the face of data changes. It is observed that the model achieved its best performance in the sudden shift scenario between Low and High workloads, with an F1-Score of 0.8719, suggesting that despite the abrupt change, the pattern of the new workload was more easily identified. On the other hand, the sudden shift from Low to Medium resulted in lower performance, with an F1-Score of 0.8111, indicating greater difficulty in adapting to the new pattern. 

The gradual shift scenario presented intermediate results with an F1-Score of 0.8525, reflecting the model's limitation in tracking progressive changes over time. Finally, the worst performance was observed in the recurring shift scenario, with an F1-Score of 0.7033 and an accuracy of only 0.5809. This result highlights the static model's fragility when faced with frequent alternations between different concepts, requiring continuous adaptation capability.

\begin{table}[h!]
\caption{Model performance under different workload shift scenarios}
\centering
\small
\renewcommand{\arraystretch}{1.1}
\begin{tabular}{lcccc}
\hline
\textbf{Type of Shift} & \textbf{Accuracy} & \textbf{Precision} & \textbf{Recall} & \textbf{F1-Score} \\
\hline
Sudden Shift (Low–Medium) & 0.7343 & 0.8521 & 0.7738 & 0.8111 \\
Sudden Shift (Low–High)   & 0.8219 & 0.9056 & 0.8406 & 0.8719 \\
Gradual Shift             & 0.7792 & 0.9220 & 0.7927 & 0.8525 \\
Recurring Shift           & 0.5809 & 0.7918 & 0.6325 & 0.7033 \\
\hline
\end{tabular}
\label{tab:resultados_rf}
\end{table}
\subsection{Application of Adaptive Detectors}

With the workload shift scenarios properly defined, adaptive detectors originally developed for concept drift were applied to evaluate the model's ability to adapt under workload shift. Specifically, we tested DDM and ADWIN, which signal a shift in data behavior based on statistical criteria and then trigger model retraining. Each detector was applied across the three simulated types of workload shift (sudden, gradual, and recurring) to assess its effectiveness in maintaining performance under dynamic conditions.

Table~\ref{tab:resultados_comparados} presents the performance of the adaptive model using the DDM detector. A significant improvement was observed in the gradual shift scenario, with an {F1-Score} of 0.9301, and a moderate recovery in the recurring shift case with 0.7131, when compared to the static model. However, in sudden shift scenarios, there was a reduction in performance. This behavior may be related to how DDM operates: since the method depends on the error rate to signal changes, it requires a minimum number of incorrect samples to statistically confirm drift. In abrupt changes, this detection time may be insufficient, and the model ends up being retrained with data still mixed between old and new concepts. This overlap compromises the model's ability to adapt to the new context.

In contrast, the ADWIN-based model showed significant improvement in all scenarios, achieving an {F1-Score} above 0.93 across all types of workload shift. ADWIN's performance is attributed to its adaptive window mechanism, which not only detects changes but also dynamically adjusts the sampling window, retaining only the most recent data and discarding obsolete information. As a result, the model is able to adapt more quickly to the new concept and maintain a high level of performance.

\begin{table}[h!]
\caption{Performance of adaptive models using DDM and ADWIN under different workload shifts}
\centering
\small
\renewcommand{\arraystretch}{1.1}
\begin{tabular}{llcccc}
\hline
\textbf{Model} & \textbf{Scenario} & \textbf{Accuracy} & \textbf{Precision} & \textbf{Recall} & \textbf{F1-Score} \\
\hline
DDM   & Sudden Shift (Low–Medium) & 0.6217 & 0.8484 & 0.5927 & 0.6979 \\
DDM   & Sudden Shift (Low–High)   & 0.8011 & 0.8562 & 0.8703 & 0.8632 \\
DDM   & Gradual Shift             & 0.8863 & 0.9195 & 0.9410 & 0.9301 \\
DDM   & Recurring Shift           & 0.6139 & 0.8558 & 0.6111 & 0.7131 \\
\hline
ADWIN & Sudden Shift (Low–Medium) & 0.9838 & 0.9921 & 0.9859 & 0.9890 \\
ADWIN & Sudden Shift (Low–High)   & 0.9141 & 0.9844 & 0.8950 & 0.9376 \\
ADWIN & Gradual Shift             & 0.9651 & 0.9859 & 0.9706 & 0.9782 \\
ADWIN & Recurring Shift           & 0.9751 & 0.9828 & 0.9855 & 0.9842 \\
\hline
\end{tabular}
\label{tab:resultados_comparados}
\end{table}

These results highlight the importance of considering workload-induced performance degradation in the development of predictive and classification models, especially for software aging detection in real-time applications. Adaptive detection mechanisms, such as ADWIN, have proven effective in mitigating the effects of workload-induced performance degradation, particularly in dynamic environments commonly observed in aging-prone systems. The ability to quickly adapt to workload shifts is fundamental to preserving model performance over time.

\subsection{Threats to Validity}

Threats to the validity of this work are described below.

\begin{itemize}
    \item \textbf{Dataset Scope.} The study is based on a single monitoring dataset collected from a SQL Server environment under controlled workload conditions. Although the dataset reflects realistic aging behavior, results may not generalize to other types of systems or resource consumption patterns.

    \item \textbf{Baseline Definition.} The baseline model was trained only under the Low workload profile to serve as a controlled reference and to highlight performance degradation when exposed to unseen workloads. Additional baselines (e.g., Medium and High) could provide complementary insights.

    \item \textbf{Synthetic Workload Transitions.} The workload shift scenarios were synthetically generated by combining segments of different workload intensities. While this design enables controlled experimentation and reproducibility, it may not fully capture the complexity of real-world workload dynamics. For example, organic variations in user demand, background processes, often introduce subtler, noisier, and less predictable shifts.

    \item \textbf{Workload Shift Simplification.} Workload shifts were modeled mainly as changes in request intensity. Although this captures a key aspect of system usage, real-world workloads may also vary in operation mix, query complexity, and temporal patterns.

    \item \textbf{Detector Configuration.} Both DDM and ADWIN rely on internal thresholds and sliding window parameters. In this study, fixed configurations were used across all experiments. Although this ensures comparability, such configurations may not be optimal for every workload shift scenario or data distribution.

  \item \textbf{Single Model Architecture.} The experiments employed a RF algorithm. While robust and interpretable, results might differ when using other model types (e.g., neural networks, ensemble techniques, or online learners). 
\end{itemize}

\section{Conclusion} \label{sec:conclusion}

Software aging represents an ongoing challenge to the reliability of long-running systems. Although ML-based detection is promising, its effectiveness is often compromised by dynamic workload and operating condition changes, which may impact model performance over time. This study investigated the robustness of software aging detection models under such workload shifts that directly affect model generalization. While these changes are not treated as classical behavior shifts, they represent operational shifts that alter data patterns over time, challenging static models and motivating the use of adaptive techniques. To this end, we compared the performance of a static baseline model with adaptive models using the DDM and ADWIN detectors under scenarios of sudden, gradual, and recurring workload shifts.

The experimental results confirmed that the static model, despite its excellent performance in a controlled environment, suffers significant degradation when exposed to new workload profiles, showing the importance of adaptability. The adaptive approach using the DDM detector presented modest improvements but was limited in scenarios involving abrupt changes. In contrast, the adaptive model with ADWIN achieved superior performance in all tested scenarios, maintaining a high F1-Score and demonstrating robustness and adaptability. The main contribution of this study is to show that adaptive strategies are not only viable but essential for effective software aging detection in dynamic production environments. Additionally, we provide a direct comparison between the DDM and ADWIN detectors, indicating that ADWIN's ability to dynamically adjust the data window is fundamental for more agile and accurate adaptation. The proposed methodology can also serve as an evaluation framework for other adaptive solutions under varying workload conditions.

As future work, we plan to investigate true concept drift scenarios, where workload and usage patterns gradually change over extended periods, progressively degrading model accuracy. In addition, we intend to extend the analysis to different types of systems, including cloud services, operating systems, and mobile devices, incorporating additional indicators such as CPU usage and response time. We also aim to explore adaptive neural network approaches and evaluate dynamic tuning of drift detector configurations to further enhance robustness across diverse operational scenarios. Another direction is to study unified or continuous aging scores instead of binary classification, as well as to design cost-aware rejuvenation strategies that balance unnecessary interventions against failure risks.

\end{document}